\documentclass[aps,prb,twocolumn,superscriptaddress,showpacs]{revtex4-1}
\usepackage{graphicx,bm,mathtools}

\begin{document}

\title{Investigations of the superconducting states of noncentrosymmetric LaPdSi$_3$ and LaPtSi$_3$}
\date{\today}

\author{M. Smidman}
\email[]{m.smidman@warwick.ac.uk}
\affiliation{Department of Physics, University of Warwick, Coventry CV4 7AL, United Kingdom}
\author{A. D. Hillier}
\email[]{adrian.hillier@stfc.ac.uk}
\affiliation{ISIS Facility, STFC, Rutherford Appleton Laboratory, Chilton, Didcot, Oxfordshire OX11 0QX, United Kingdom}
\author{D. T. Adroja}
\affiliation{ISIS Facility, STFC, Rutherford Appleton Laboratory, Chilton, Didcot, Oxfordshire OX11 0QX, United Kingdom}
\affiliation{Physics Department, University of Johannesburg, P.O. Box 524, Auckland Park 2006, South Africa}
\author{M. R. Lees}
\affiliation{Department of Physics, University of Warwick, Coventry CV4 7AL, United Kingdom}
\author{V. K. Anand}
\affiliation{ISIS Facility, STFC, Rutherford Appleton Laboratory, Chilton, Didcot, Oxfordshire OX11 0QX, United Kingdom}
\author{R. P. Singh}
\affiliation{Department of Physics, University of Warwick, Coventry CV4 7AL, United Kingdom}
\author{R. I. Smith}
\affiliation{ISIS Facility, STFC, Rutherford Appleton Laboratory, Chilton, Didcot, Oxfordshire OX11 0QX, United Kingdom}
\author{D. M. Paul}
\affiliation{Department of Physics, University of Warwick, Coventry CV4 7AL, United Kingdom}
\author{G. Balakrishnan}
\email[]{g.balakrishnan@warwick.ac.uk}
\affiliation{Department of Physics, University of Warwick, Coventry CV4 7AL, United Kingdom}

\date{\today}

\begin{abstract}
The noncentrosymmetric superconductors LaPdSi$_3$ and LaPtSi$_3$ have been studied with magnetization, specific heat, resistivity and $\mu$SR measurements. These crystallize in the tetragonal BaNiSn$_3$ structure and superconductivity is observed at  $T_{\rm{c}}~=~2.65(5)$~K for LaPdSi$_3$ and  $T_{\rm{c}}~=~1.52(6)~$K for  LaPtSi$_3$. The results are consistent with both compounds being weakly coupled, fully gapped superconductors but $\mu$SR measurements reveal that LaPdSi$_3$ is a bulk type-I superconductor while LaPtSi$_3$ is a type-II material with a Ginzburg-Landau parameter of $\kappa~=~2.49(4)$. This is further supported by specific heat measurements, where the transition in an applied field is first-order in LaPdSi$_3$ but second-order in LaPtSi$_3$. The electronic specific heat in the superconducting state was analyzed using an isotropic $s$-wave model that gave $\Delta_0/k_{\rm B}T_{\rm{c}}~=~1.757(4)$ for LaPdSi$_3$ and 1.735(5) for LaPtSi$_3$. The temperature dependence of the effective penetration depth($\lambda_{\rm{eff}}(T)$) of LaPtSi$_3$ was extracted from $\mu$SR measurements and was fitted giving $\Delta_0/k_{\rm B}T_{\rm{c}}~=~1.60(8)$ and $\lambda_{\rm{eff}}(0)~=~239(3)$~nm. A critical field of $B_{\rm{c}}$(0)~=~182.7~G was obtained for LaPdSi$_3$ from $\mu$SR measurements, which is in good agreement with the calculated thermodynamic critical field. 
\end{abstract}


\maketitle

\section{Introduction}
There has been considerable recent interest in studying noncentrosymmetric superconductors (NCS),\cite{BauerNCS} which do not have a center of inversion in their crystal structure. A lack of inversion symmetry along with a finite antisymmetric spin-orbit interaction means that the superconducting states may no longer be entirely spin-singlet or spin-triplet, but an admixture of the two.\cite{NCSGorkov} Interest in NCS was initially triggered by the discovery of the coexistence of antiferromagnetic order ($T_{\rm{N}}$~=~2.2~K) and superconductivity ($T_{\rm{c}}$~=~0.75~K) in the heavy-fermion compound CePt$_3$Si (space group $P4mm$).\cite{CePt3Si2004} Pressure induced superconductivity was subsequently observed in the noncentrosymmetric antiferromagnets CeRhSi$_3$ ,\cite{CeRhSi3SC} CeIrSi$_3$, \cite{CeIrSi3SC} CeCoGe$_3$ \cite{CeCoGe3SC} and CeIrGe$_3$, \cite{CeIrGe3SC} all with the space group $I4mm$. All these cerium-based NCS crystallize in a tetragonal structure in which a lack of a mirror plane perpendicular to [001] leads to a Rashba type antisymmetric spin-orbit coupling.\cite{BauerNCS} In addition to the difficulty in experimentally accessing the superconducting states of many of these compounds, the effects of strong electronic correlations make it difficult to discern the role of inversion symmetry in determining the nature of the superconductivity. As a result, NCS without strong correlations have been increasingly studied. Some of these compounds such as Li$_2$Pt$_3$B \cite{Li2Pt3BRep,Li2Pt3BNode}, LaNiC$_2$ \cite{LaNiC2Rep,LaNiC2TRS}, Re$_6$Zr \cite{Re6Zr} and the locally NCS SrPtAs \cite{SrPtAs} have been reported to display unconventional properties. Others such as Li$_2$Pd$_3$B  \cite{Li2Pd3BRep,Li2Pt3BNode}, $T_2$Ga$_9$~($T$~=~Rh,~Ir) \cite{T2Ga9Rep,T2Ga92009} and Nb$_{0.18}$Re$_{0.82}$ \cite{NbRe2011} appear to behave as conventional, $s$-wave superconductors. There has been particular interest in systems where the spin-orbit coupling can be varied by the substitution of atoms of a different mass. For example, the aforementioned isostructural NCS Li$_2$Pd$_3$B and Li$_2$Pt$_3$B were reported to display a change from fully gapped to nodal superconductivity upon the substitution of Pt for Pd.\cite{Li2Pt3BNode} However, there are a limited number of NCS reported to display unconventional behavior. It is therefore important to find further systems where there is evidence for gap structures resulting from singlet-triplet mixing  and to examine the effects of varying the spin-orbit coupling. 

There have also been several recent studies of NCS of the form $RTX_3$ ($R$~=~La, Ba, Sr, $T$~=~transition metal, $X$~=~Si or Ge) with the tetragonal BaNiSn$_3$ structure (space group $I4mm$), the same as the cerium based, pressure-induced NCS.\cite{BaPtSi32009,CaMSi32011,RTX32011,LaRhSi3Rep,LaRhSi32011,LaPdSi3Rep} Of these, LaRhSi$_3$ is a type-I superconductor \cite{LaRhSi32011} whereas BaPtSi$_3$ is a type-II material with a relatively low upper critical field of $B_{\rm{c2}}$(0)~=~640~G. \cite{RTX32011} Type-I superconductivity has also been observed in the NCS $T_2$Ga$_9$ \cite{T2Ga92009} and specific heat measurements of LaPt$_3$Si indicate a first-order transition in field \cite{LaPt3Si2011}, as expected for a type-I material. Bulk superconductivity has been reported in LaPdSi$_3$ from specific heat measurements with $T_{\rm{c}}$~=~2.6~K,\cite{LaPdSi3Rep} while no superconductivity was reported in LaPtSi$_3$ down to 2~K.\cite{EuPtSi32010} We report magnetic susceptibility, specific heat and muon spin rotation/relaxation ($\mu$SR) measurements on LaPdSi$_3$ and LaPtSi$_3$ and demonstrate that LaPtSi$_3$ is a type-II superconductor with $T_{\rm{c}}$~=~1.52~K, whereas LaPdSi$_3$ is a type-I compound. 

\section{Experimental}
Polycrystalline samples of LaPdSi$_3$ and LaPtSi$_3$ were prepared by arc-melting stoichiometric quantities of the constituent elements on a water cooled copper hearth, in an argon atmosphere. The samples were flipped and melted several times to improve homogeneity and wrapped in Ta foil, sealed in evacuated quartz tubes and annealed at 900~$^{\circ}$C for two weeks. Powder x-ray diffraction measurements were carried out using a Panalytical X-Pert Pro diffractometer. Magnetic susceptibility measurements were made using a Magnetic Property Measurement System superconducting quantum interference device magnetometer (Quantum Design). Measurements between 0.48 and 1.8~K were made using an iQuantum $^3$He insert. Specific heat measurements were made between 0.4 and 3~K using the two-tau relaxation method with a Quantum Design Physical Properties Measurement System (PPMS) with a $^3$He insert. Neutron scattering and $\mu$SR measurements were carried out at the ISIS facility at the Rutherford Appleton Laboratory, UK. Neutron diffraction data were collected at room temperature using the General Materials Diffractometer (GEM).\cite{GEM2005} Approximately 12~g of the sample was placed in a 6~mm diameter, thin-walled cylindrical vanadium can and data collected in all six detector banks were simultaneously fitted. $\mu$SR measurements were performed using the MuSR spectrometer, with detectors in both the longitudinal and transverse geometry. Spin-polarized muons were implanted into the sample. In the longitudinal configuration, the positrons were detected either in forward or backward positions along the axis of the muon beam. The asymmetry is calculated by
\begin{equation}
G_z(t) = \frac{N_F - \alpha N_B}{N_F + \alpha N_B},
\end{equation}
where $N_F$ and $N_B$ are the number of counts at the detectors in the forward and backward positions and $\alpha$ is determined from calibration measurements taken with a small applied transverse magnetic field. In this configuration measurements were made in zero field with an active compensation system which cancel stray fields to within 0.01~G. In the transverse configuration, a field was applied perpendicular to the direction of the muon beam and the detectors were grouped into two orthogonal pairs. For one grouping, the detectors are in the forward and backward positions, for the other they are in the top and bottom. For the transverse field measurements of LaPtSi$_3$, the spectra from the two detector sets were simultaneously fitted while for the measurements of LaPdSi$_3$, only the spectra from the top and bottom detector grouping were used.

Throughout this work the CGS system of electromagnetic units have been used. Applied fields have been denoted by $H$ in units of Oe and internal fields by $B$ in units of G.

\section{Results and Discussion}

\begin{figure}[tb]
   \includegraphics[width=0.99\columnwidth]{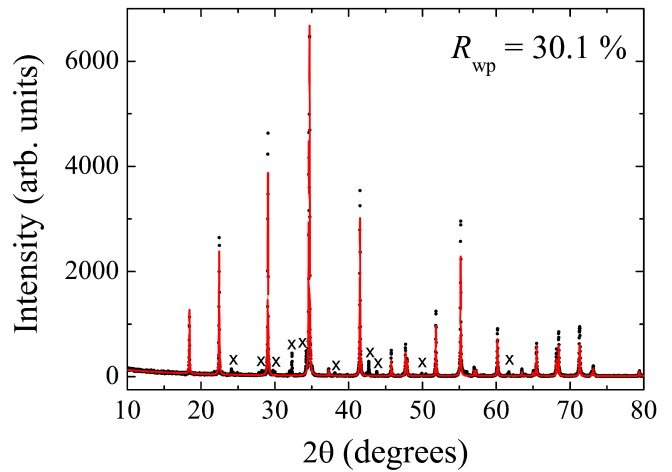}
	\caption{Fitted  powder x-ray diffraction measurements of LaPdSi$_3$ at room temperature. A Rietveld refinement is indicated by solid lines and the crosses indicate impurity peaks. The weighted profile factor ($R_{\rm{wp}}$) is shown.}
   \label{LaPdSi3XRD}
\end{figure}

\begin{figure}[tb]
   \includegraphics[width=0.99\columnwidth]{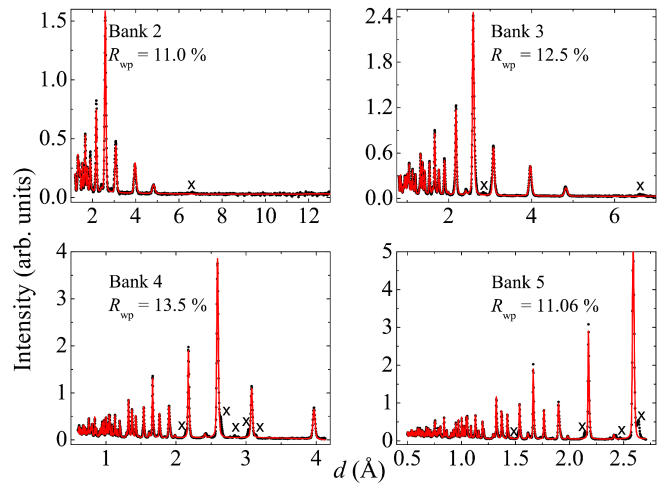}
	\caption{Fitted powder neutron diffraction profiles in selected GEM detector banks from LaPtSi$_3$ at room temperature. A Rietveld refinement is indicated by solid lines and the crosses indicate impurity peaks.  The weighted profile factors ($R_{\rm{wp}}$) for each bank are shown}
   \label{LaPtSi3ND}
\end{figure}

\subsection{Structural studies}

The crystal structures of the title compounds were refined by the Rietveld method in the tetragonal space group $I4mm$ (number 107) using either powder x-ray (LaPdSi$_3$) or neutron (LaPtSi$_3$) diffraction data, with the co-ordinates of the isostructural BaNiSn$_3$ used in the structural model. The x-ray diffraction data were fitted using the TOPAS academic software \cite{TOPAS} (Fig.~\ref{LaPdSi3XRD}), while the neutron data were fitted using the General Structure Analysis System (GSAS) \cite{ExpGui} (Fig.~\ref{LaPtSi3ND}). Refined unit cell parameters were $a~=~4.3542(4)$ and $c~=~9.664(1)~\rm{\AA}$ for LaPdSi$_3$; and $a~=~4.3474(2)$ and $c~=~9.6368(6)~\rm{\AA}$ were obtained for LaPtSi$_3$, which  are in good agreement with previously reported values. \cite{LaPdSi3Rep,EuPtSi32010} A small number of impurity peaks were observed in both compounds. The first and third most intense unfitted peaks in LaPdSi$_3$ are consistent with an impurity phase of LaSi$_2$ with the orthorhombic $\alpha$-GdSi$_2$ structure with a weight fraction smaller than 5\%. The impurity peaks for LaPtSi$_3$ could not be indexed to any La-Pt-Si compounds in the 2013 ICDD Powder Diffraction File.\cite{ICDD}

\subsection{LaPdSi$_3$}

\begin{figure}[tb]
   \includegraphics[width=0.99\columnwidth]{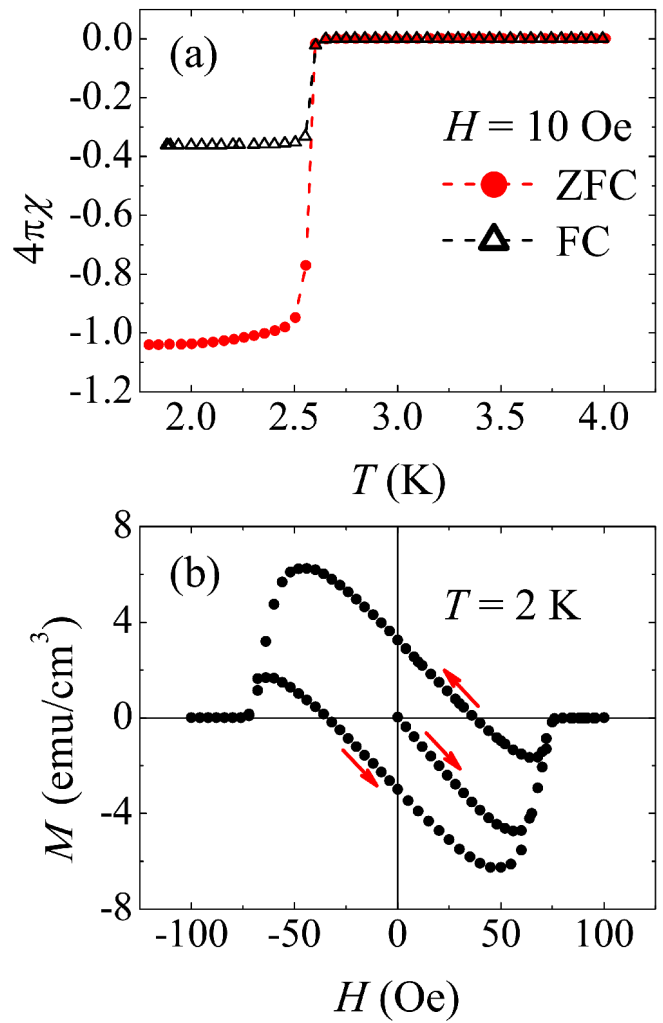}
	\caption{(a) Magnetic susceptibility of LaPdSi$_3$ in an applied field of 10~Oe. Both zero-field cooled (ZFC) and field-cooled (FC) measurements are shown. Corrections for demagnetization effects have been made after Ref.~\onlinecite{DeMag}. (b) Magnetization as a function of applied field measured at 2~K. }
   \label{LaPdSi3_MvT_MvH}
\end{figure}

Magnetic susceptibility ($\chi$) measurements of LaPdSi$_3$ as a function of temperature were carried out down to 1.8~K in an applied field of 10~Oe. As shown in Fig.~\ref{LaPdSi3_MvT_MvH}(a), a sharp superconducting transition is observed at 2.6~K. After correcting for demagnetizing effects using the expressions in Ref.~\onlinecite{DeMag}, the zero-field cooled (ZFC) volume susceptibility curve falls to around $4\pi$$\chi~\sim~-1$, indicating complete flux expulsion and bulk superconductivity in the compound. The magnetization as a function of applied field at 2~K is shown in Fig.~\ref{LaPdSi3_MvT_MvH}(b). The low field value of $\frac{dM}{dH}$ of the virgin curve once again indicates the complete expulsion of magnetic fields from the sample. The abrupt change in gradient of the magnetization and loss of diamagnetism at an applied field of 80~Oe indicates that bulk superconductivity has been suppressed. However, at fields lower than this there is a region where the magnetization is reversible. Upon decreasing the field from 100~Oe, there is a partial recovery of diamagnetism as magnetic flux is expelled from the sample. This behavior is very different from the magnetization curves often observed in type-II superconductors but is expected for type-I superconductors. The reentrance of diamagnetism upon decreasing the applied field has also been observed in the type-I materials LaRhSi$_3$ and $T_2$Ga$_9$. \cite{LaRhSi32011,T2Ga92009} In an ideal type-I superconductor, the magnetization is linear as a function of field until the critical field ($H_{\rm{c}}$) where there is a discontinuous jump to the normal state behavior. However, the effect of a finite demagnetization factor is to broaden the transition and the system enters the intermediate state where there are macroscopic normal and superconducting domains.\cite{WaldramSC}

\begin{figure}[tb]
   \includegraphics[width=0.99\columnwidth]{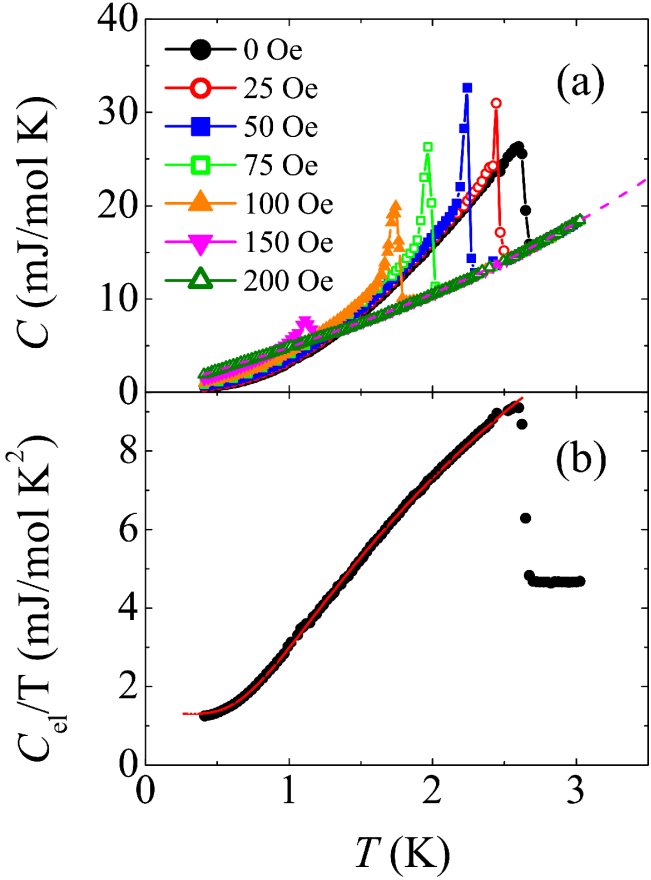}
	\caption{(a)Specific heat of LaPdSi$_3$ in zero and applied fields up to 200~Oe. The dashed line shows a fit to the normal state described in the text. (b)Electronic contribution to the specific heat in zero field, obtained from subtracting an estimate of the phonon contribution. The solid line shows a fit to a BCS model described in the text.}
   \label{LaPdSi3_CvT}
\end{figure}

\begin{figure}[tb]
   \includegraphics[width=0.99\columnwidth]{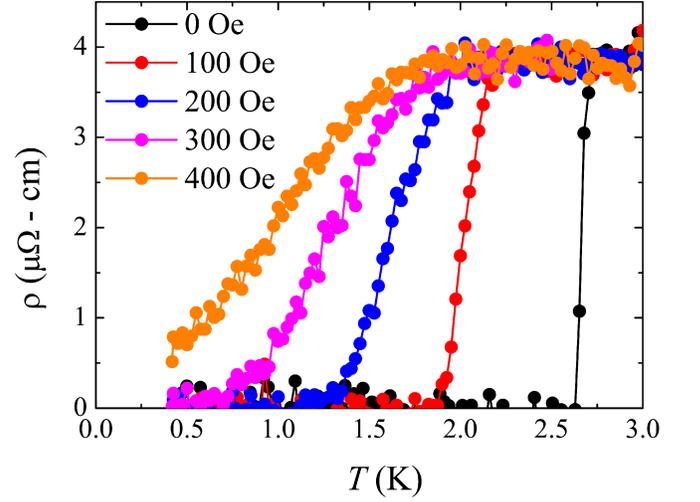}
	\caption{Temperature dependence of the resistivity of LaPdSi$_3$ across the superconducting transition, in applied fields up to 400~Oe.}
   \label{LaPdSi3RvTFields}
\end{figure}

The specific heat in zero-field and applied fields up to 200~Oe are shown in Fig.~\ref{LaPdSi3_CvT}(a). A jump in the specific heat in zero-field indicates the onset of bulk superconductivity. Defining the transition temperature as the midpoint of the transition gives $T_{\rm{c}}~=~(2.65~\pm~0.05)~$K. The dashed line shows a fit to the normal state of $C_{\rm{n}}$($T$)$~=~\gamma T~+~\beta T^3$ which yields $\gamma~=~4.67(4)$~mJ/mol~K$^2$ and  $\beta~=~0.155(5)$~mJ/mol~K$^4$.  The Debye temperature can be calculated using $\theta_{\rm{D}}~=~(12\pi^4N_{\rm{A}}nk_{\rm{B}}/5\beta)^{\frac{1}{3}}$ and $\theta_{\rm{D}}~=~397(4)$~K is obtained. A calculation of the electron-phonon coupling constant ($\lambda_{\rm{e-ph}}$) following Ref.~\onlinecite{McMill} gives $\lambda_{\rm{e-ph}}~=~0.51$, putting LaPdSi$_3$ in the weak coupling limit. With the application of a magnetic field,  $T_{\rm{c}}$ is rapidly suppressed and bulk superconductivity is not observed above 0.4~K with an applied field of 200~Oe. As well as rapidly suppressing superconductivity, the transitions sharpen when measured in an applied field. The jump in the specific heat at the transition is also larger in applied fields of 25, 50 and 75~Oe than it is with zero applied field. This suggests that the superconducting transition is second-order in zero applied field but first-order in an applied field. This is further evidence that LaPdSi$_3$ is a type-I superconductor.\cite{WaldramSC}

The temperature dependence of the electronic contribution ($C_{\rm{el}}/T$) is shown in Fig.~\ref{LaPdSi3_CvT}(b), obtained from $C_{\rm{el}}~=~C~-~\beta T^3$. The offset of $C_{\rm{el}}/T$ from zero at low temperatures indicates the presence of non superconducting fraction, most likely resulting from the presence of impurity phases. The data were fitted to a BCS model of the specific heat. The entropy ($S$) was calculated from

\begin{equation}
\frac{S}{\gamma T_{\rm{c}}}~=~-\frac{6}{\pi^2}\frac{\Delta_0}{k_{\rm B}T_{\rm{c}}} \int_0^{\infty}[f {\rm{ln}} f + (1~-~f ){\rm{ln}} (1~-~f)] dy,
\label{Entropy}
\end{equation}

\noindent where f is the Fermi-Dirac function given by $f~=~(1~+~e^{E/k_{\rm{B}}T})^{-1}$ and E~=~$\Delta_0\sqrt{y^2+ \delta(T)^2}$ where $y$ is the energy of the normal state electrons and $\delta(T)$ is the temperature dependence of the superconducting gap calculated from BCS theory. Both quantities have been normalized by the magnitude of the gap at zero temperature ($\Delta_0$). The specific heat of the superconducting state is calculated by

\begin{equation}
\frac{C_{\rm{sc}}}{\gamma T}~=~\frac{\rm{d} (S/\gamma T_{\rm{c}})}{\rm{d}t}.
\label{specheat}
\end{equation}

\noindent This method reproduces the dataset in Ref.~\onlinecite{BCSDat} for the value of  $\Delta_0/k_{\rm B}T_{\rm{c}}$ from BCS theory. The data were fitted by scaling equation~\ref{specheat} and adding a constant background from the non-superconducting fraction. The scaling corresponds to either a reduced molar superconducting fraction ($a_{\rm{sc}}$) or a value of  $\gamma$ of the superconducting state ($\gamma_{\rm{sc}}$) different to the value measured in the normal state of $\gamma~=~4.67(4)$~mJ/mol~K$^2$. The data were fitted with $a_{\rm sc}\gamma_{\rm sc}~=~3.366(11)~$mJ/mol~K$^2$.  Using  $a_{\rm{sc}}~=~0.765$ from the estimate of the  superconducting volume fraction from $\mu$SR measurements, $\gamma_{\rm{sc}}$~=~4.40(1)~mJ/mol~K$^2$ is obtained. $\Delta_0/k_{\rm B}T_{\rm{c}}~=~1.757(4)$ was fitted which is very close to the BCS value of 1.764. \cite{WaldramSC} $\Delta$$C/\gamma$$T_{\rm{c}}$ obtained using the normal state $\gamma$ is $\sim~0.99$ but with the fitted parameter of $a_{\rm sc}\gamma_{\rm sc}$, the value is $\sim~1.37$ which is close to the BCS value.

The temperature dependence of the resistivity across the superconducting transition in several fields up to 400~Oe is shown in Fig.~\ref{LaPdSi3RvTFields}. In zero-field a sharp transition is observed with an onset at $T^{\rm{onset}}_{\rm{c}}~=~2.70(3)~$K and zero resistivity at $T^{\rm{zero}}_{\rm{c}}~=~2.63(3)~$K. Upon the application of a magnetic field, $T^{\rm{zero}}_{\rm{c}}$ is strongly suppressed but the temperature of the superconducting onset is not greatly reduced, resulting in a significant broadening of the transition.

\begin{figure}[tb]
   \includegraphics[width=0.99\columnwidth]{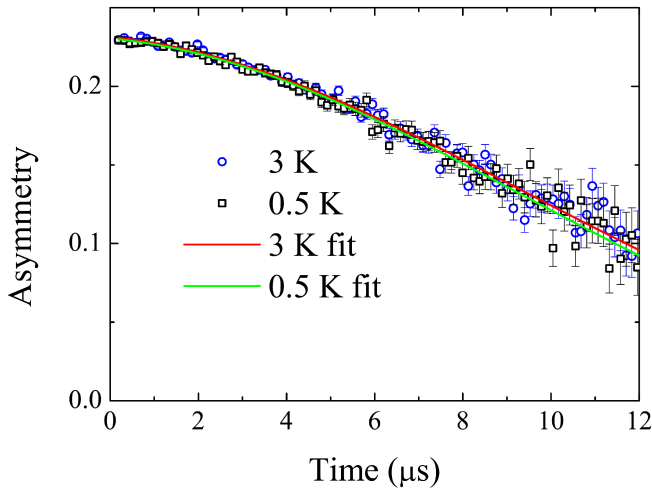}
	\caption{Zero-field $\mu$SR spectra at 0.5 and 3~K. The solid lines show fits to the Kubo-Toyabe function described in the text.}
   \label{LaPdSi3ZF}
\end{figure}

Muon spin relaxation/rotation measurements were also used to study the superconducting state of LaPdSi$_3$. Fig.~\ref{LaPdSi3ZF} shows the $\mu$SR spectra measured in zero field at 0.5 and 3~K. These were fit with a Kubo-Toyabe function

\begin{equation}
G_z(t) = A \left[ \frac{1}{3} + \frac{2}{3} ( 1 - \sigma^2 t^2 ) \rm{exp} \left( - \frac{\sigma^2 t^2}{2}\right)\right] \rm{exp} (-\Lambda t),
\label{KTEq}
\end{equation}

\noindent where $A$ is the initial asymmetry, $\sigma$ is the static relaxation rate and $\Lambda$ is the electronic relaxation rate. $\sigma$ is a measure of the distribution of fields which are static on the timescale of the muon lifetime ($\sim$2.2~$\mu$s) while $\Lambda$ measures the distribution of fluctuating fields. Generally the main contribution to $\sigma$ is from nuclear moments whilst electronic moments contribute to $\Lambda$. In systems where the superconducting state breaks time reversal symmetry (TRS), spontaneous magnetic moments arise below $T_{\rm{c}}$ and an increase may be observed in either $\sigma$ \cite{LaNiGa22012} or $\Lambda$.\cite{LaNiC2TRS} At 3~K, $\sigma~=~0.0692(14)~\mu\rm{s}^{-1}$ and $\Lambda~=~0.012(2)~\mu\rm{s}^{-1}$ were obtained and $\sigma~=~0.071(2)~\mu\rm{s}^{-1}$ and $\Lambda~=~0.011(2)~\mu\rm{s}^{-1}$ were obtained at 0.5~K. Therefore no evidence of TRS breaking is observed in LaPdSi$_3$.

\begin{figure}[tb]
   \includegraphics[width=0.99\columnwidth]{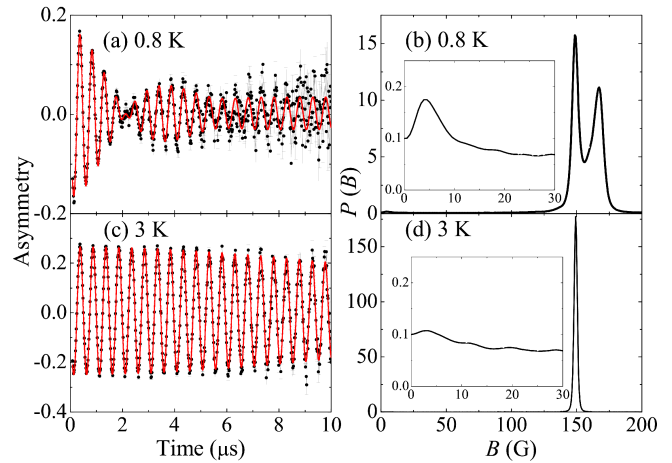}
	\caption{ Transverse field $\mu$SR spectra of LaPdSi$_3$ at (a) 0.8~K and (c) 3~K for an applied field of 150~Oe. Panels (b) and (d) show the maximum entropy spectra for the respective temperatures. The insets show the maximum entropy spectra at low values of $B$.}
   \label{LaPdSi3TF}
\end{figure}

Transverse field $\mu$SR measurements were made with the sample field cooled in applied fields up to 300~Oe. Figures~\ref{LaPdSi3TF}(a) and \ref{LaPdSi3TF}(c) show the spectra at 0.8 and 3~K, below and above $T_{\rm{c}}$. Below $T_{\rm{c}}$, the depolarization rate of the asymmetry sharply increases, indicating bulk superconductivity in the sample. There is also a reduction in the initial asymmetry upon entering the superconducting state. The maximum entropy spectra which show the magnetic field probability distribution ($P(B)$) are shown in Fig.~\ref{LaPdSi3TF}(b) and \ref{LaPdSi3TF}(d). At 3~K a sharp peak is observed at 150~G. At 0.8~K the peak at the applied field broadens and an additional peak is present at a field greater than the applied field. This generally had an asymmetric profile with a longer tail in the low field direction. The presence of an internal field at a greater frequency than the applied field is strong evidence for bulk type-I superconductivity in the compound. For an applied field of $B~<~B_{\rm{c}}$, demagnetization effects may mean that some regions of the superconductor have a field applied greater than $B_{\rm{c}}$, in which case magnetic flux can penetrate the bulk of the sample. Muons implanted in these normal regions of the intermediate state will precess at a frequency corresponding to the field at the muon site which must be at least equal to $B_{\rm{c}}$. Muons implanted in regions where magnetic flux is expelled will only be affected by nuclear moments. This accounts for the peak present at low fields in Fig.~\ref{LaPdSi3TF}(b) but absent in Fig.~\ref{LaPdSi3TF}(d). 

\begin{figure}[tb]
   \includegraphics[width=0.99\columnwidth]{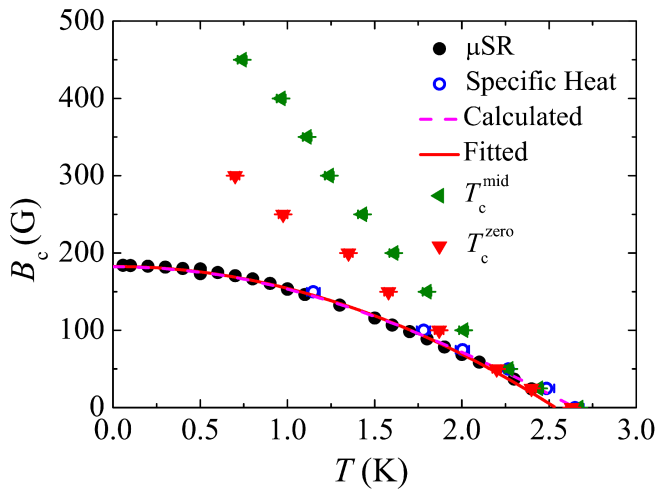}
	\caption{Temperature dependence of the critical field ($B_{\rm{c}}$) of LaPdSi$_3$ obtained from specific heat, $\mu$SR and resistivity measurements. The data labelled $\mu$SR were obtained from fitting the time spectra using Equation \ref{TFFit}. The field which is larger than the applied field has been taken to be $B_{\rm{c}}$. The solid line shows a fit to equation \ref{HcFit} and the dashed line is a calculation of the thermodynamic critical field using equation~\ref{HcCalc}.}
   \label{LaPdSi3_MuRes}
\end{figure}

The asymmetries were fit to the expression

\begin{equation}
G_z(t) = \sum_{i=1}^{n} A_i {\rm{cos(\gamma_\mu}} B_i t+\phi)e^{-(\sigma_i t)^2/2} + A_{\rm{bg}},
\label{TFFit}
\end{equation}

\noindent where $A_{\rm{i}}$ are the amplitudes of the oscillatory component precessing about a magnetic field $B_{\rm{i}}$ with a Gaussian decay rate $\sigma_{\rm{i}}$. There is a common phase $\phi$, a background term  $A_{\rm{bg}}$ and $\rm{{\gamma_\mu}}/{2\pi}$~=~135.53 MHz~T$^{-1}$. Three oscillatory components were fitted for all of the LaPdSi$_3$ spectra. Fixing $B_2$~=~$B_3$ and $\sigma_3$~=~0 means that there are two components ($A_2$ and $A_3$) precessing about the applied field, one with a decaying component and one not. With $A_2~=~0.0528$ in the superconducting state and a total asymmetry from the sample in the normal state of 0.225, the non-superconducting volume fraction is estimated to be 23.5~\%. Since $B_1$ is greater than the applied field, it is taken to be equal to $B_{\rm{c}}$. The temperature dependence of  $B_{\rm{c}}$ is shown in Fig.~\ref{LaPdSi3_MuRes} and the values obtained from $\mu$SR are in good agreement with those obtained from the specific heat. In particular, there is not a significant change in $B_1$ for different applied fields. This can be seen from the fact several measurements were taken at 0.5~K in applied fields from 50 to 160~Oe and there is good agreement between the obtained values of $B_{\rm{c}}$. The critical field was fit to the expression

\begin{equation}
B_{\rm{c}}(T) = B_{\rm{c}}(0) \left[1 - \left(\frac{T}{T_{\rm{c}}}\right)^2\right].
\label{HcFit}
\end{equation}

\noindent Values of $B_{\rm{c}}$(0)~=~182.7(7)~G and $T_{\rm{c}}$~=~2.54(1)~K were obtained. The critical field is slightly higher than that observed in the isostructural LaRhSi$_3$. \cite{LaRhSi32011} The dashed line in Fig.~\ref{LaPdSi3_MuRes} shows a calculation of $B_{\rm{c}}$ obtained from calculating the difference between the free energies per unit volume the normal and superconducting states ($\Delta$$F$) by

\begin{equation}
\frac{B_{\rm{c}}^2(T)}{8\pi}~=\Delta F~=~\int_{T_{\rm{c}}}^{T} \int_{T_{\rm{c}}}^{T'} \frac{C_{\rm{sc}}-C_{\rm{n}}}{T''}dT''dT',
\label{HcCalc}
\end{equation}

\noindent where $C_{\rm{n}}$ and $C_{\rm{sc}}$ are the heat capacities per unit volume. $B_{\rm{c}}(T)$ was calculated using $\gamma_{\rm{sc}}$~=~4.40~mJ/mol~K$^2$  and $\Delta_0/k_{\rm B}T_{\rm{c}}~=~1.757(4)$. This is in good agreement with the data, with a calculated value of $B_{\rm{c}}$(0)~=~182.1~G. Also displayed in Fig.~\ref{LaPdSi3_MuRes} are the critical field values from resistivity measurements, obtained from defining $T_{\rm{c}}$ as the temperature where $\rho~=~0$ and at the midpoint of the transition. These results indicate the presence of surface superconductivity with critical fields above that of the bulk values. The points taken from the midpoint of the transition also show a positive curvature, turning up at low temperatures.

\subsection{LaPtSi$_3$}

\begin{figure}[tb]
   \includegraphics[width=0.99\columnwidth]{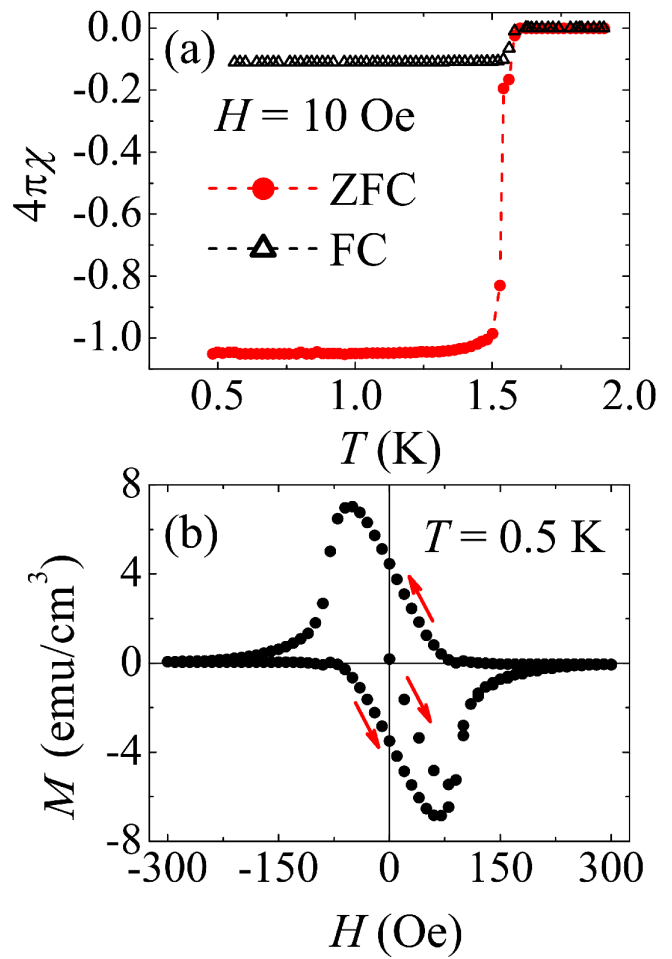}
	\caption{(a) Temperature dependence of the magnetic susceptibility of LaPtSi$_3$. The sample was cooled in zero field before being measured in an applied field of 10~Oe.(b) Magnetization as a function of applied field at 0.5~K.}
   \label{LaPtSi3MvT}
\end{figure}
\begin{figure}[tb]
   \includegraphics[width=0.99\columnwidth]{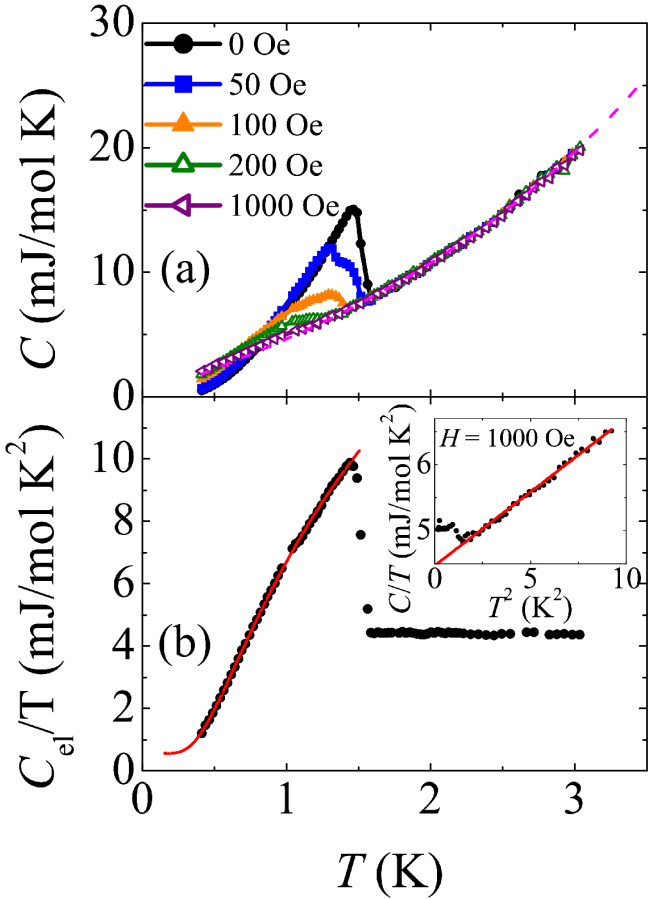}
	\caption{(a)Specific heat of LaPtSi$_3$ in zero and applied fields up to 1000~Oe. The dashed line shows a fit to the normal state described in the text. (b) Electronic contribution to the specific heat in zero field, obtained from subtracting an estimate of the phonon contribution. The solid line shows a fit to a BCS model described in the text. The inset shows $C/T$ vs $T^2$ for an applied field of 1000~Oe.}
   \label{LaPtSi3CvT}
\end{figure}

\begin{figure}[tb]
   \includegraphics[width=0.99\columnwidth]{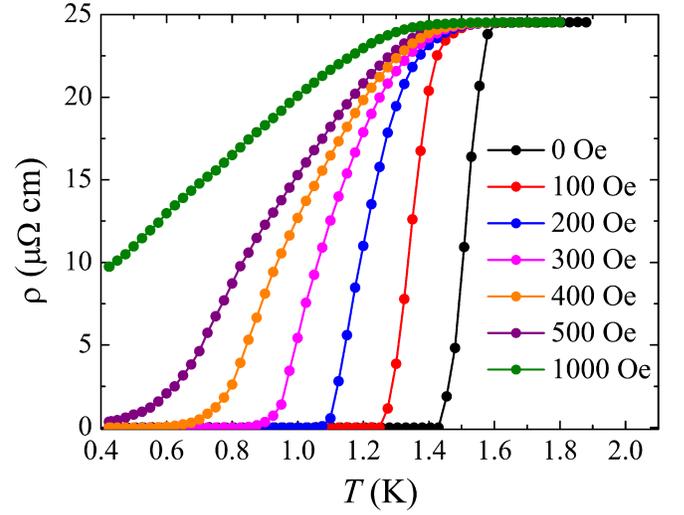}
	\caption{Temperature dependence of the resistivity of LaPtSi$_3$ across the superconducting transition, measured down to 0.4~K.}
   \label{RvTFields}
\end{figure}

Figure~\ref{LaPtSi3MvT}(a) shows the temperature dependence of the magnetic susceptibility of LaPtSi$_3$ down to 0.48~K in an applied magnetic field of 10~Oe. A sharp superconducting transition is observed with an onset at 1.58~K. After correcting for demagnetization effects, $4\pi$$\chi~\sim$~-1 is obtained indicating bulk superconductivity in the compound. The zero-field specific heat (Fig.~\ref{LaPtSi3CvT}(a)) shows a bulk superconducting transition with $T_{\rm{c}}~=~1.52(6)~$K. The in-field measurements show significant broadening compared to those taken in zero field, particularly with applied fields of 100 and 200~Oe. The normal state data in zero field was fit above $T_{\rm{c}}$ with $C_{\rm{n}}$($T$)$~=~\gamma$$T~+~\beta$$T^3$ with $\gamma~=~4.41(4)$~mJ/mol~K$^2$ and  $\beta~=~0.238(5)$~mJ/mol~K$^4$, giving $\theta_{\rm{D}}~=~344(2)$~K. This is a similar value to that of LaRhSi$_3$ \cite{LaRhSi32011} and BaPtSi$_3$ \cite{BaPtSi32009} and $\lambda_{\rm{e-ph}}~\sim~0.47$ is calculated, putting LaPtSi$_3$ in the weak coupling limit. As well as becoming broader in field, the jump in the specific heat at the transition is smaller than in zero-field. This suggests that the superconducting transition is second-order in field, indicating type-II superconductivity. The inset of Fig.~\ref{LaPtSi3CvT}(b) shows $C/T$ against $T^2$ in an applied field of 1000~Oe. The line shows a fit to the normal state. At around $1.1~$K there is a deviation from linear behavior which may correspond to a superconducting transition with a significantly reduced volume fraction. As shown in Fig.~\ref{RvTFields} a superconducting transition is observed in the resistivity in this applied field, which onsets at a similar temperature. The electronic contribution to the specific heat in zero-field (Fig.~\ref{LaPtSi3CvT}(b)) was fitted by scaling equation \ref{specheat} with a constant background to account for the non-superconducting fraction. A good fit to the data was obtained which corresponds to a molar superconducting fraction of $a_{\rm{sc}}$~=~93\% for $\gamma_{\rm{sc}}~=~4.41$~mJ/mol~K$^2$. $\Delta_0/k_{\rm B}T_{\rm{c}}~=~1.735(5)$ was obtained from the fit which is slightly below the BCS value. The observed jump in the specific heat is $\Delta$$C/\gamma$$T_{\rm{c}}~\sim~1.33$ and the BCS value of $\sim1.43$ is obtained, taking into account the observed $a_{\rm{sc}}$. 

There is further evidence for type-II superconductivity from the magnetization as a function of applied field at 0.5~K (Fig.~\ref{LaPtSi3MvT}(b)). The form of the hysteresis loop strongly resembles that expected for a type-II superconductor in the presence of flux pinning.\cite{WaldramSC} Unlike LaPdSi$_3$, upon reducing the applied field, there is no reentrance of diamagnetism, distinguishing this from type-I behavior.

\begin{figure}[tb]
   \includegraphics[width=0.99\columnwidth]{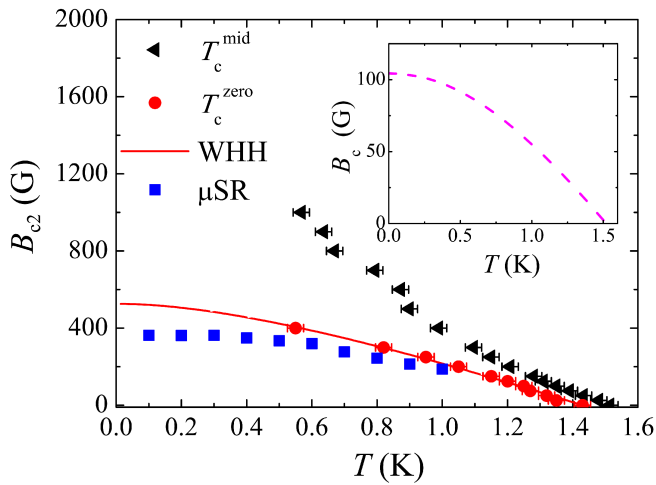}
	\caption{Temperature dependence of the upper critical field obtained from resistivity and $\mu$SR measurements. The solid line shows a fit made to the latter with a WHH model as described in the text. The blue points show the bulk values of $B_{\rm{c}2}$ obtained  from $\mu$SR measurements. A calculation of the critical field using equation~\ref{HcCalc} is shown in the inset.}
   \label{LaPtSi3Hc2}
\end{figure}

The temperature dependence of the resistivity across the superconducting transition in several fields up to 1000~Oe is shown in Fig.~\ref{RvTFields}. The transition in zero-field is sharp, with an onset at $T^{\rm{onset}}_{\rm{c}}~=~1.58(2)~$K and zero resistivity at $T^{\rm{zero}}_{\rm{c}}~=~1.43(2)~$K. As with LaPdSi$_3$, there is a significant broadening of the transition in field as  $T^{\rm{zero}}_{\rm{c}}$ is suppressed much more rapidly than  $T^{\rm{onset}}_{\rm{c}}$. Fig.~\ref{LaPtSi3Hc2} shows the temperature dependence of $B_{\rm{c2}}$ obtained from the midpoint of the resistive transition ($T^{\rm{mid}}_{\rm{c}}$), $T^{\rm{zero}}_{\rm{c}}$ and $\mu$SR measurements. The latter correspond to values of $B_{\rm{c2}}$ for the sample bulk and these measurements are described further on in this section. The $B_{\rm{c2}}$ data corresponding to $T^{\rm{zero}}_{\rm{c}}$ are slightly larger than those of the bulk but this is likely to be because resistivity measurements measure the surface superconductivity which is expected to be more robust. The data were fitted using the Werthamer-Helfand-Hohenberg (WHH) model in the dirty limit.\cite{WHH3} In fitting the model, the expression $\alpha~=~5.2758\times10^{-5}(-$d$B_{\rm{c2}}$/d$T)_{T~=~T_{\rm{c}}}$ was used, where d$B_{\rm{c2}}$/d$T$ is units of G/K and $\alpha$ is the Maki parameter. From the fit to the WHH model, $\alpha~=0.0280(3)$ is obtained. Since $\alpha$ is proportional to the ratio of $B_{\rm{c2}}$ and the Pauli paramagnetic limiting field ($\mu$$H_{\rm{p}}$), the low value of $\alpha$ implies that orbital pair breaking is the dominant mechanism for destroying superconductivity. This is expected, since  $\mu$$H_{\rm{p}}$ is calculated to be 28.3~kG using $T_{\rm{c}}~=1.52$~K. Since this is much greater than $B_{\rm{c2}}(0)$~=~526~G, the effect of paramagnetic pair breaking will be small. The WHH model also contains an additional parameter which measures the degree of spin-orbit coupling ($\lambda_{\rm{so}}$). However at low values of the $\alpha$, the WHH model is insensitive to changes in $\lambda_{\rm{so}}$, so this parameter was fixed to zero when fitting. The values of $B_{\rm{c2}}$ obtained from $T^{\rm{mid}}_{\rm{c}}$ show a positive curvature down to 0.4~K. This is unlike those obtained from  $T^{\rm{zero}}_{\rm{c}}$ and demonstrates the significant broadening of the transition in field. A positive curvature of $B_{\rm{c2}}$ from resistivity measurements is also observed in BaPtSi$_3$, CaPtSi$_3$ and CaIrSi$_3$.\cite{BaPtSi32009,CaMSi32011} The inset of Fig.~\ref{LaPtSi3Hc2} shows a calculation of the thermodynamic critical field using the parameters obtained from fitting the electronic specific heat with  $B_{\rm{c}}$(0)~=~104.3~G being obtained.

\begin{figure}[tb]
   \includegraphics[width=0.99\columnwidth]{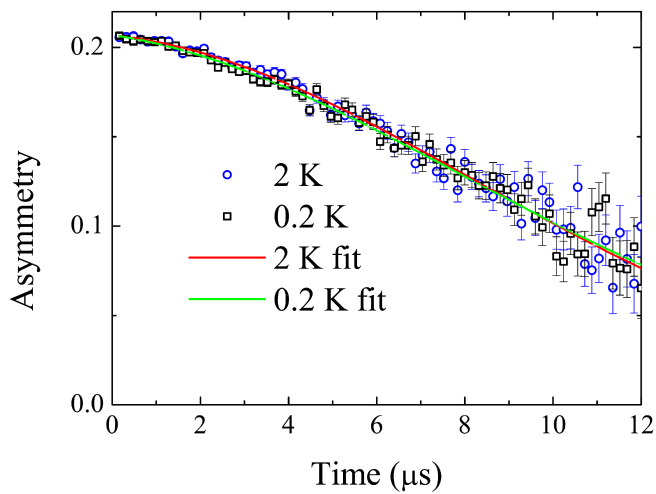}
	\caption{Zero-field $\mu$SR spectra of LaPtSi$_3$ at 0.2 and 2~K. The solid lines show fits to the Kubo-Toyabe function described in the text.}
   \label{LaPtSi3ZF}
\end{figure}

\begin{figure}[tb]
   \includegraphics[width=0.99\columnwidth]{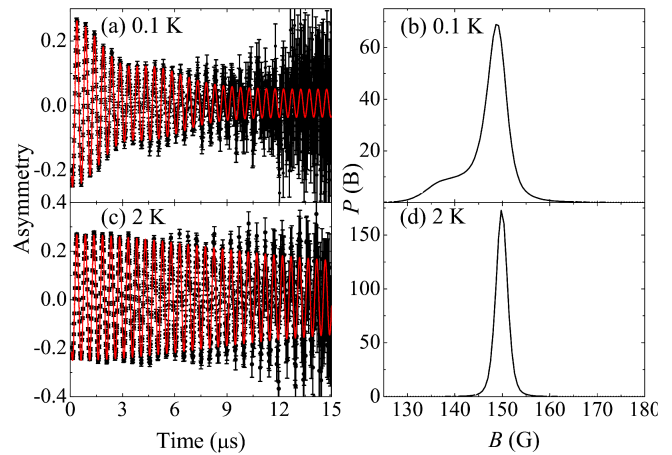}
	\caption{Transverse $\mu$SR spectra of LaPtSi$_3$ at (a) 0.1 and (c) 2~K, in an applied field of 150~Oe. The corresponding maximum entropy spectra are shown in (b) and (d).}
   \label{LaPtSi3TF}
\end{figure}

$\mu$SR measurements were carried out on LaPtSi$_3$. Zero-field measurements are shown above and below $T_{\rm{c}}$ in Fig.~\ref{LaPtSi3ZF} with fits made to equation~\ref{KTEq}. $\sigma~=~0.079(1)~\mu\rm{s}^{-1}$ and $\Lambda~=~0.013(2)~\mu\rm{s}^{-1}$ were obtained at 2~K and  $\sigma~=~0.078(2)~\mu\rm{s}^{-1}$ and $\Lambda~=~0.014(2)~\mu\rm{s}^{-1}$ were obtained at 0.2~K. Therefore there is no evidence that TRS is broken in the superconducting state of LaPtSi$_3$. Transverse field $\mu$SR measurements were carried out in several applied fields up to 400~Oe. The spectra at 0.1 and 2~K in an applied transverse field of 150~Oe are shown in Figs.~\ref{LaPtSi3TF}(a) and \ref{LaPtSi3TF}(c) respectively. A significant increase in the depolarization rate upon entering the superconducting state indicates the onset of bulk superconductivity. The corresponding maximum entropy spectra are shown in Figs.~\ref{LaPtSi3TF}(b) and \ref{LaPtSi3TF}(d). In the normal state the spectra show a peak in $P(B)$ centered around an applied field. In the superconducting state the peak around the applied field broadens and an additional shoulder in the distribution is observed at lower fields. This is very different to the field distribution observed in the superconducting state of LaPdSi$_3$ (Fig.~\ref{LaPdSi3TF}) and indicates bulk type-II superconductivity. This is the field distribution of the flux-line lattice in the mixed state, where most of the contribution to $P(B)$ is at fields less than the applied field.

\begin{figure}[tb]
   \includegraphics[width=0.99\columnwidth]{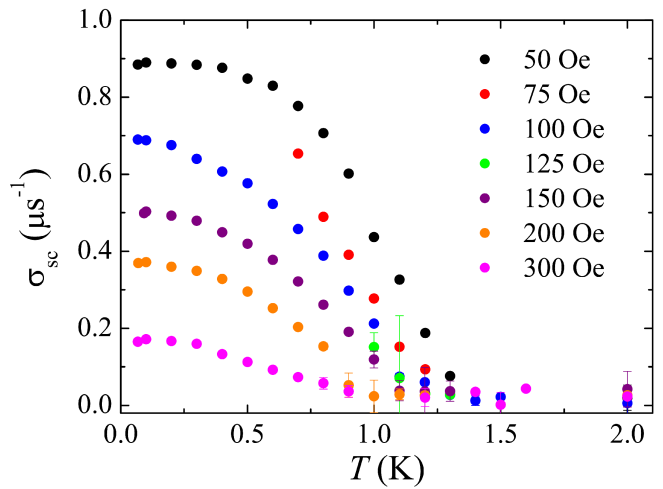}
	\caption{Temperature dependence of the superconducting contribution to the variance of LaPtSi$_3$ for applied fields of 50 to 300~Oe.}
   \label{sigscT}
\end{figure}

\begin{figure}[tb]
   \includegraphics[width=0.99\columnwidth]{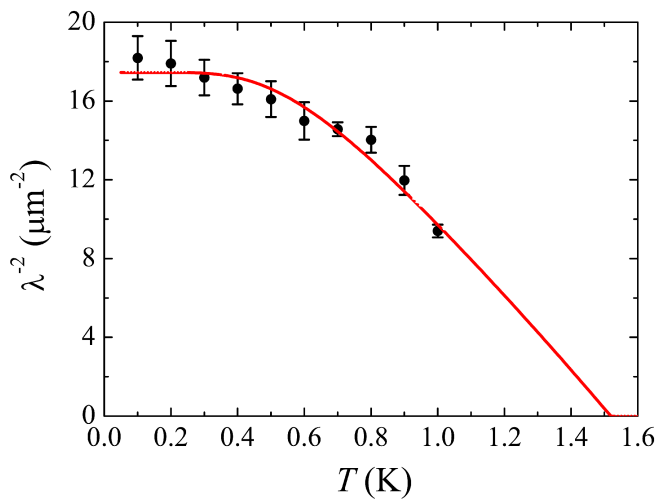}
	\caption{Temperature dependence of the London penetration depth of LaPtSi$_3$. The solid line shows a fit to Equation~\ref{LamTEq}.}
   \label{LamT}
\end{figure}

The spectra were fit using equation \ref{TFFit} with  $n~=~3$ in the superconducting state except for an applied field of 300~Oe which could only be fitted with $n~=~2$. Once again $\sigma_3$~=~0, and the component $A_3$ corresponds to those muons implanted in the silver sample holder. The temperature dependence of the effective penetration depth ($\lambda_{\rm{eff}}$($T$)) and $B_{\rm{c2}}$ were obtained following the multiple Gaussian method outlined in Ref.~\onlinecite{MultiGauss}. The first and second moments of the field distribution in the sample were calculated by

\begin{equation}
 \langle B \rangle= \sum_{i=1}^{n - 1} \frac{A_i B_i}{A_{\rm{tot}}},
\label{FirstMom}
\end{equation}

\begin{equation}
 \langle B^2 \rangle= \sum_{i=1}^{n - 1} \frac{A_i}{A_{\rm{tot}}}[(\sigma_i/\gamma_\mu)^2 + (B_i -  \langle B \rangle)^2],
\label{SecMom}
\end{equation}

\noindent where $A_i$ are the weighting factors and $A_{\rm{tot}}~=~ \sum_{i=1}^{n - 1}A_i$. The superconducting contribution to the second moment ($\langle B^2 \rangle_{\rm{sc}}$) is obtained from subtracting the second moment obtained in the normal state. The superconducting component of the variance ($\sigma_{\rm{sc}}$) is proportional to $\sqrt{\langle B^2 \rangle_{\rm{sc}}}$ and the temperature dependence in several applied fields is shown in Fig.~\ref{sigscT}. For applied fields of $B~\ll~B_{\rm c2}$ and large values of the Ginzburg-Landau parameter ($\kappa$), $\sigma_{\rm sc}$ is field independent and is proportional to $\lambda_{\rm{eff}}^{-2}$. Evidently this is not the case for LaPtSi$_3$ and therefore the field dependence of $\sigma_{\rm{sc}}$ must be modeled. At each temperature the following expression was fitted

\begin{equation}
\sigma_{\rm{sc}}~=4.83~\times 10^4 \frac{\kappa^2 (1 - b)}{\lambda_{\rm{eff}}^2 (\kappa^2 - 0.069)},
\label{sigsc_b}
\end{equation}

\noindent where $\sigma_{\rm{sc}}$ is in units of $\mu$s$^{-1}$, $\lambda_{\rm{eff}}$ is in nm, $b~=~\mu$$H/B_{\rm{c2}}$ and $H$ is the applied field. This expression was used in Ref.~\onlinecite{CaC6} for the low $\kappa$ superconductor CaC$_6$ and is based on an approximation given in Ref.~\onlinecite{Brandt2003}, valid for all $\kappa$ for $b~>~0.25$. Using $\kappa~=~\lambda_{\rm{eff}}/\xi$ and $B_{\rm{c2}}~=~\Phi_0/2\pi\xi^2$, where $\phi_0$ is the magnetic flux quantum and $\xi$ is the Ginzburg-Landau coherence length, equation~\ref{sigsc_b} reduces to two free parameters and was fitted to isotherms from 0.1 to 1~K. The temperature dependence of the obtained values of $B_{\rm{c2}}$ are shown in Fig.~\ref{LaPtSi3Hc2}. The curve is slightly lower than that obtained from resistivity but this is expected since $\mu$SR is a bulk probe rather than measuring the superconductivity of the surface. The temperature dependence of the penetration depth is shown in Fig.~\ref{LamT}. $\lambda_{\rm{eff}}^{-2}$ is proportional to the superfluid density and could be fitted to an isotropic, $s$-wave model as follows

\begin{equation}
\frac{\lambda_{\rm{eff}}^{-2}(T)}{\lambda_{\rm{eff}}^{-2}(0)}~=~1 + \frac{1}{\pi} \int_0^{2\pi} \int_{\Delta(T)}^\infty \frac{\partial f}{\partial E} \frac{E dE d\phi}{\sqrt{E^2 - \Delta(T)}},
\label{LamTEq}
\end{equation}

\noindent where $f$ is the Fermi function, $\lambda_{\rm{eff}}(0)$ is the effective penetration depth at zero temperature and $\Delta(T)$ is the superconducting gap. For an isotropic, $s$-wave model, $\Delta(T)$ has no dependence on $\phi$ and was approximated by $\Delta(T)~=~\Delta_0\tanh(1.82(1.018(T_{\rm{c}}/T-1))^{0.51})$, where $\Delta_0$ is the gap magnitude at zero temperature.\cite{CarringtonMgB2} Fixing $T_{\rm{c}}$ to the value observed in the specific heat, $\Delta_0$~=~0.209(7)~meV was obtained, giving $\Delta_0/k_{\rm B}T_{\rm{c}}~=~1.60(8)$. This is slightly below the BCS value of 1.764 and the data are compatible with a fully gapped, weakly coupled superconductor.  $\lambda_{\rm{eff}}(0)~=~239(3)$~nm was obtained from the fit and using $\xi~=~96(1)$~nm from $B_{\rm{c2}}(0)$, $\kappa~=~2.49(4)$ is obtained. This indicates LaPtSi$_3$ is in the low $\kappa$ regime. This is in good agreement with the value of 2.44(7) calculated using  $\kappa~=~B_{\rm{c2}}/\sqrt{2}B_{\rm{c}}$.\cite{Tinkham}  A value of $\kappa~<~5$ validates the use of equation~\ref{sigsc_b}, rather than the more commonly used expression where $\sigma_{\rm{sc}}~\propto~(1-b)(1+1.21(1~-\sqrt{b})^3)$. This equation could in fact be fitted to the data and very similar results were obtained for $\lambda_{\rm{eff}}$ and $B_{\rm{c2}}(0)$.

\begin{table}[ht]
\caption{Superconducting parameters of LaPdSi$_3$ and LaPtSi$_3$.}
\label{SCTab} 
\begin{ruledtabular}
 \begin{tabular}{c c c }

&  LaPdSi$_3$ & LaPtSi$_3$ \\
\hline
$T_{\rm{c}}$ (K)&2.65(5)&1.52(6) \\
$\lambda_{\rm{e-ph}}$ &0.51& 0.47 \\
$\Delta_0/k_{\rm B}T_{\rm{c}}$&1.757(4)&1.735(5) - specific heat \\
&&1.60(8) - $\mu$SR\\
$\lambda_{\rm{eff}}$(nm)&&239(3)\\
$\xi$(nm)&&96(1)\\
$\kappa$&&2.49(4)\\
$B_{\rm{c}}$(0)~(G)&182.1 - calculated &104.3 - calculated\\
&182.7 - $\mu$SR&\\
$B_{\rm{c2}}$(0)~(G)&&360(10) - $\mu$SR\\
&&526 - resistivity ($\rho~=~0)$\\

\end{tabular}
\end{ruledtabular}
\end{table}

\section{Summary}

We have studied the noncentrosymmetric superconductors LaPdSi$_3$ and LaPtSi$_3$. Various superconducting parameters determined from our measurements are shown in Table~\ref{SCTab}. Magnetization, specific heat and $\mu$SR measurements reveal that LaPdSi$_3$ is a bulk type-I superconductor. The specific heat measurements reveal that the superconducting transition is second-order in zero-field but first-order in an applied field, as expected for a type-I superconductor. $\mu$SR measurements confirm the presence of bulk type-I superconductivity. With an applied transverse field, a fraction of muons are implanted in an environment with a local magnetic field larger than the applied field. This is consistent with probing macroscopic normal regions of the intermediate state. The critical field ($B_{\rm{c}}$) is deduced from the value of this field and is in excellent agreement with those measured from the specific heat measurements and $B_{\rm{c}}(0)$~=~182.7(7)~G is obtained from the analysis of the temperature dependence of $B_{\rm{c}}(T)$. This is in good agreement with the calculated thermodynamic critical field. The electronic contribution to the specific heat was fitted with $\Delta_0/k_{\rm B}T_{\rm{c}}~=~1.757(4)$ which is very close to the BCS value. We also report that LaPtSi$_3$ is a bulk superconductor with $T_{\rm{c}}$~=~1.52(6)~K. In contrast to LaPdSi$_3$, magnetization, specific heat and $\mu$SR measurements reveal type-II superconductivity in LaPtSi$_3$. Specific heat measurements reveal that the superconducting transition is second-order in both zero and applied fields. $\mu$SR measurements are used to probe the field distribution of the mixed state and the temperature dependence of $\lambda_{\rm eff}$ and $B_{\rm{c2}}(0)$ are obtained from the field dependence of the second moment of magnetization. Zero temperature values of $\lambda_{\rm eff}(0)$~=~239(3)~nm and $\xi(0)$~=~96(1)~nm give $\kappa~=~2.49(4)$. An isotropic, $s$-wave model was fitted to $\lambda_{\rm eff}$ giving $\Delta_0/k_{\rm B}T_{\rm{c}}~=~1.60(8)$ while a value of 1.735(5) was obtained from fitting the specific heat. The specific heat of LaPdSi$_3$ at low temperatures is consistent with a fully gapped, $s$-wave model while the specific heat of LaPtSi$_3$ is not measured to low enough temperatures to confirm the gapped behavior. The $\mu$SR data could be fitted to a single fully-gapped model. However this does not exclude the possibility of two gaps since in NCS with a small triplet component, both gaps may be nodeless and also of a similar magnitude. \citep{NCSswave} Therefore, more detailed measurements may be able to discern the presence of two gaps. The jump in the specific heat at the transition $\Delta$$C/\gamma$$T_{\rm{c}}$ is lower in both compounds than the value predicted by BCS theory but this may largely be accounted for by the reduced superconducting fraction of the samples. The bulk values of $B_{\rm{c2}}(0)$ are compared to the values obtained from resistivity measurements. The resistive transition significantly broadens in field and $B_{\rm{c2}}(T)$ deduced from the midpoint of the transition shows positive curvature down to 0.4~K. In the WHH model, the low value of $\alpha$ indicates that orbital pair breaking is the dominant pair breaking mechanism and the effect of Pauli paramagnetic limiting is negligible.  

The effect of substituting Pt for Pd in the LaTSi$_3$ system is to increase $\kappa$ and drive the system from type-I to type-II behavior. In the dirty limit, $\kappa~=~0.715\lambda_{\rm L}/l$ where $\lambda_{\rm L}$ is the London penetration depth and $l$ is the mean free path \citep{Tinkham}. A larger value of $\kappa$ may arise because of a smaller mean free path in LaPtSi$_3$ and this is supported by a larger residual resistivity of $\sim24.5~\mu\Omega$-cm compared to  $\sim3.9~\mu\Omega$-cm in LaPdSi$_3$. LaRhSi$_3$ has also been reported to be a type-I superconductor while BaPtSi$_3$ is type-II. It remains to be seen if it is a general feature of the NCS $RT\rm{Si}_3$ compounds that those where $T$ has a $4d$ outer shell are type-I and those with a $5d$ configuration are type-II. 

\begin{acknowledgments}
We acknowledge the EPSRC, UK for providing funding (grant number EP/I007210/1). DTA/ADH/VKA thank CMPC-STFC (grant number CMPC-09108) for financial support We thank T.E. Orton for technical support. Neutron (GEM Xpress Access) and muon beamtime was provided by the UK Science and Technology Facilities Council. Some of the equipment used in this research at the University of Warwick was obtained through the Science City Advanced Materials: Creating and Characterising Next Generation Advanced Materials Project, with support from Advantage West Midlands (AWM) and part funded by the European Regional Development Fund (ERDF). 
\end{acknowledgments}

\bibliography{LaTSi3_bib}

\end{document}